\documentclass[aps,prd,preprint,groupedaddress,showpacs,showkeys]{revtex4-1}
\usepackage{graphicx}
\usepackage{amssymb}
\usepackage {amsmath}
\DeclareGraphicsExtensions{.jpg}
\DeclareMathOperator{\eint}{{\mathrm E}_1}

\begin{document}


\title{Newtonian black holes: Particle production, ``Hawking" temperature, entropies and entropy field equations}


\author{Aldo Mart\' inez-Merino}
\email{a.merino@fisica.ugto.mx}
\author{Octavio Obreg\' on}
\email{octavio@fisica.ugto.mx}
\affiliation{Departamento de F\' isica,
Divisi\' on de Ciencias e Ingengier\' ias, Campus Le\' on,
Universidad de Guanajuato,
Loma del Bosque 103, Fracc. Lomas del Campestre, Le\' on, Guanajuato, M\' exico}
\author{Michael P. Ryan, Jr.}
\email{ryanmex2002@yahoo.com}
\thanks{Permanent address: 30101 Clipper Lane, Millington MD 21651, USA}
\affiliation{Instituto de Ciencias Nucleares,
Universidad Nacional Aut\' onoma de M\' exico,
A. Postal 70-543, M\' exico D.F., M\' exico}



\date{\today}

\begin{abstract}
Newtonian gravitation with some slight modifications, along with some highly simplified ideas from quantum field theory allow us to reproduce, at least at the level of back-of-the-envelope calculations, many results of black hole physics.  We consider particle production by a black hole, the Newtonian equivalent of the Hawking temperature, and the Bekenstein entropy.  Also, we are able to deduce Newtonian field equations from entropy.  We finally study higher-order Newtonian theories under the same assumptions used for ordinary Newtonian theory.  In a companion article we will look at entropic forces for various entropies and make contact with our analysis of higher-order Newtonian theories.
\end{abstract}


\maketitle

\section{Introduction}

It is interesting to note that, at least at the level of back-of-the-envelope calculations, a slightly modified Newtonian theory of gravity can give results surprisingly close to some of those in general relativity.  One does need to give the speed of light a privileged position as a sort of ``speed limit'' attached to ordinary Newtonian gravity. Of course, since Newtonian gravity is an action-at-a-distance theory, it is difficult to use it to model such things as gravitational radiation.  This said, much can be done.  Since Newtonian gravity is studied at a relatively elementary stage in physics education, the results from this modified Newtonian theory are sometimes more transparent than similar results in GR.

A classic example of this is the prediction of ``dark stars'' by Michell \cite{mich} and Laplace \cite{lap} in the 18th century, where the calculation for radial motion away from a mass $M$ of a radius where escape velocity is the speed of light, so light itself cannot escape, making any massive body smaller than this radius invisible or ``black.''
The radius of the dark star is well known to be
\begin{equation}
r = \frac{2GM}{c^2}, \label{dsrad}
\end{equation}
exactly the same as the Schwarzschild  radius in GR.  It is also known that this exact result is a coincidence.  However, one might expect from dimensional grounds something like this result but with a coefficient different from two.

\begin{figure}[h]
\includegraphics[width=8cm,height=5cm]{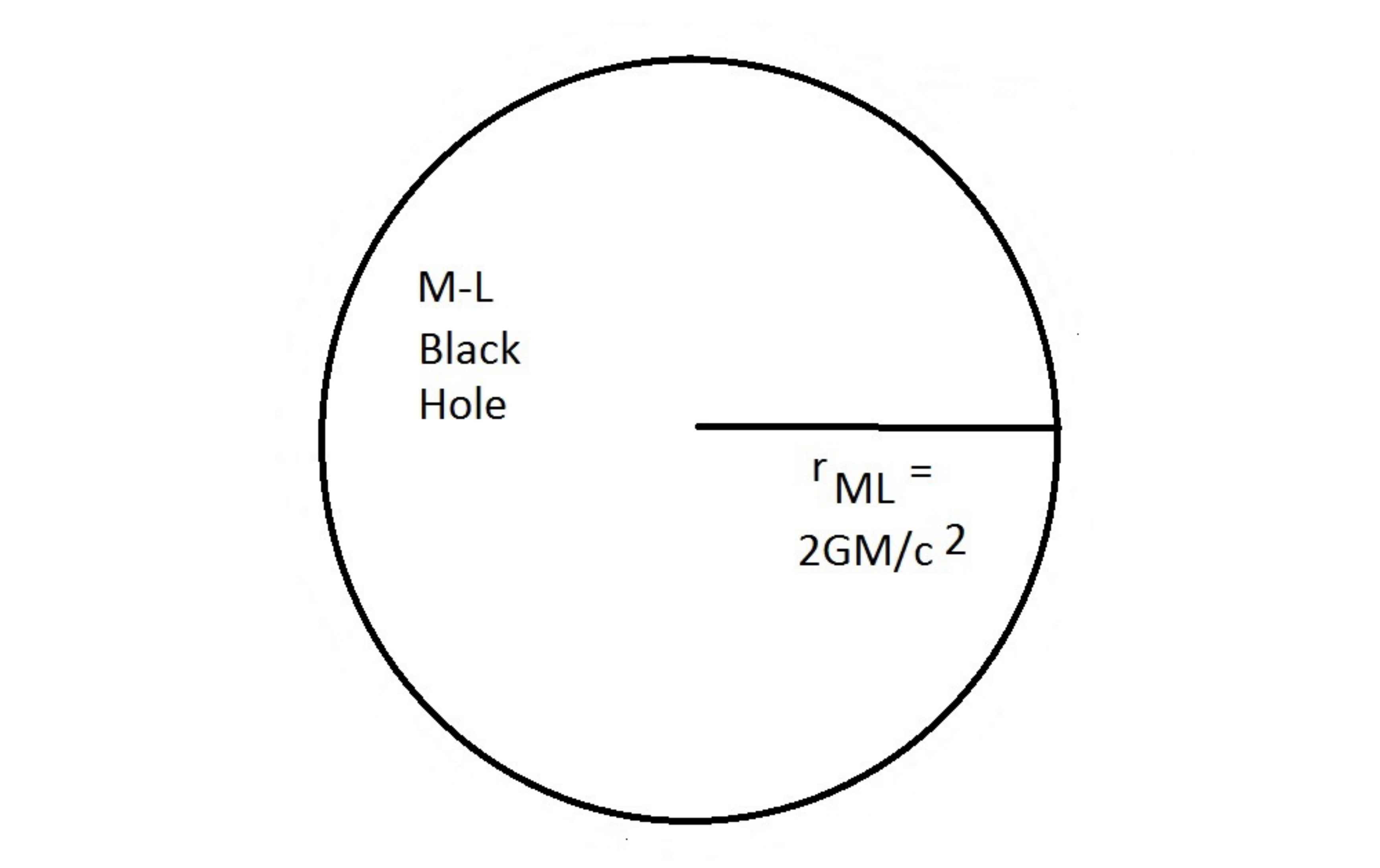}
\caption{Michell-Laplace black hole}
\label{fig:MLBH}
\end{figure}

If we now apply our ``speed limit'' of $c$, the orbital mechanics of a particle of mass $m$ will be complicated, but one thing is certain; once the particle falls below the above radius it can no longer escape, giving the Michell-Laplace dark star some important attributes of a black hole. For this reason we will call such a ``dark star'' a Michell-Laplace black hole, and the radius of the black hole the Michell-Laplace radius, $r_{ML}$.  Figure ~\ref{fig:MLBH} shows this.

Once we have the concept of a Michell-Laplace black hole, there are a number of possible calculations that can be performed that are similar to ones done in GR, with the advantage that they become much simpler and possibly more obvious.

In this simplified picture we want to study several scenarios known in GR.  In Section II the first concept we will treat is simplified particle production by the black hole and compare the results to those from a similar GR calculation discussed in the Appendix.  Section III will be devoted to a Newtonian version of Jacobson's derivation of the Einstein field equations from black hole entropy. Section IV will study higher-order Newtonian theory (a MoND theory) in order to investigate the possibility of a calculation similar to that of Guedens, Jacobson and Sarkar \cite{gued} for a higher-order relativistic theory of gravity. Section V will be conclusions and suggestions for further research.

A final note is that we have chosen to use ordinary cgs units as in Eq. (\ref{dsrad}) instead of GR units where $G = c = 1$ (and sometimes $\hbar = k_B =1$, $k_B$ the Boltzmann constant).

\section{``Hawking'' temperature and black hole entropy}

\begin{figure}[htbp]
\centering
\includegraphics[width=8cm,height=5cm]{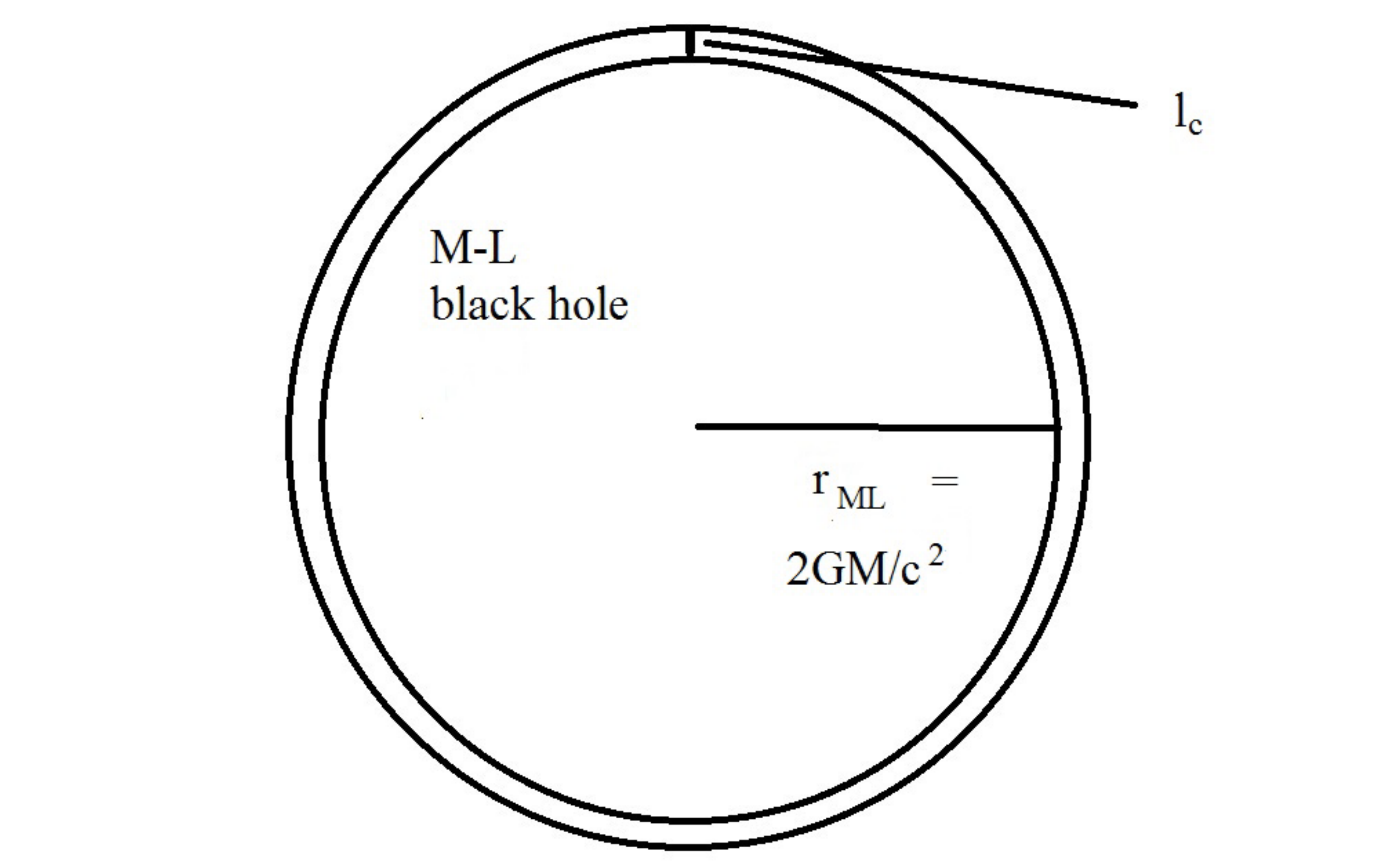}
\caption{A shell of width $l_c$ around a Michell-Laplace black hole}
\label{fig:newshell}
\end{figure}

In the Appendix we use a simplified picture of particle production by a GR black hole to calculate back-of-the-envelope values for the equivalent of the Hawking temperature by assuming that virtual production of a particle pair occurs in a shell the width of the Compton wavelength of a particle of mass $m$ around the horizon of a GR black hole. In the Newtonian case we can do the same for a Michell-Laplace black hole with the shell surrounding the surface where escape velocity is $c$.
Figure ~\ref{fig:newshell} shows this scenario.

\begin{figure}[htbp]
\centering
\includegraphics[width=8cm,height=5cm]{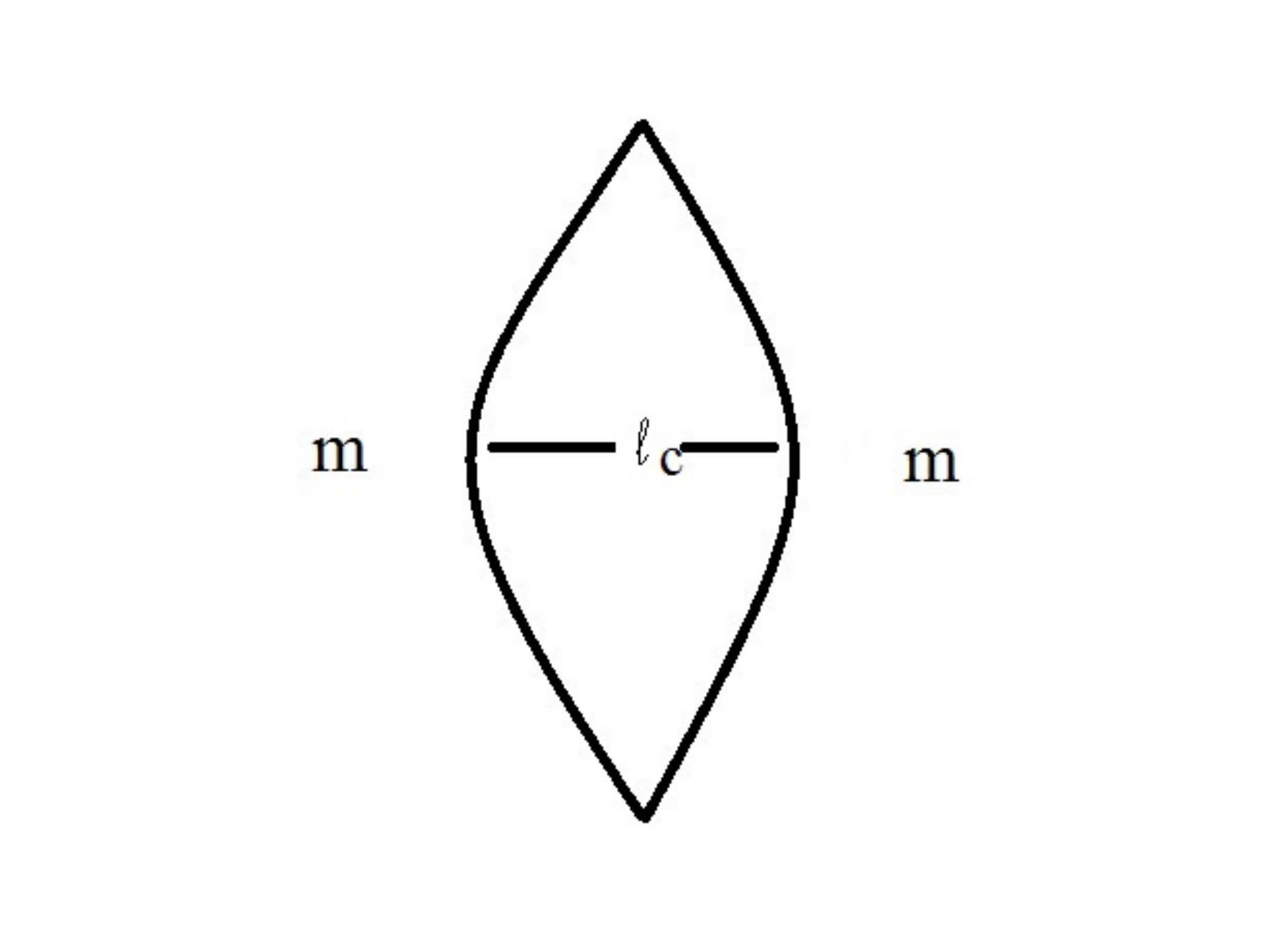}
\caption{A ``Feymnan'' diagram for the virtual production of two particles of mass $m$ }
\label{fig:gfeyn}
\end{figure}

In the Appendix, we model virtual gravitational particle production on the production of electron-positron pairs by taking the Feynman diagram in Fig. 6 there literally and assume that a strong enough electric field would split the virtual pair and produce real particles. The equivalent figure here would be  a ``Feynman'' diagram (treating the gravitational field as similar to the electromagnetic field) shown in Fig. ~\ref{fig:gfeyn}
Notice that special relativity implies that we do not expect conservation of mass as we do for conservation of charge in electromagnetism, so we can produce two particles of the same positive mass.

However, if both masses are positive, when one falls into the black hole the mass $M$ of the black hole increases, so the energy carried away by the escaping particle does not lead to black hole evaporation.  We can formally assume that one of the masses is negative, and the negative mass, for some unknown reason, {\it always} falls into the hole.  A simple calculation shows that the equation for black hole evaporation would be essentially the same as the one given by Hawking.

At a radius of $r_{ML} + l_c/2$ one of these particles could fall into the ML black hole and disappear (since it cannot exceed the ``speed limit'' $c$) and the other could escape to infinity.  We will assume that the particles will be produced with velocity $c$, an assumption we will justify shortly. We also assume that the new real particle will depart from the point of production radially away from the ML black hole.  There is no real justification for this except for the reasonableness of the final answer. 

In elementary Newtonian theory the energy per unit mass of a particle of mass $m$ in such radial motion is
\begin{equation}
\frac{E}{m} = \frac{1}{2} \left( \frac{dr}{dt} \right)^2 - \frac{GM}{r}.
\end{equation}
Now, if the particle is produced with velocity $c$ at $r_{ML} + l_c/2$,
\begin{eqnarray}
\frac{E}{m}\bigg|_{r = \frac{2GM}{c^2} + l_c/2} &=& \frac{1}{2} c^2 - \frac{GM}{r_{ML} + l_c/2} \\
&\approx& \frac{1}{2} c^2 - \frac{GM}{r_{ML}}\left ( 1 - \frac{l_c}{2r_{ML}}\right ),
\end{eqnarray}
this is the same at $r = \infty$,
\begin{eqnarray}
E_{\infty} &=& \frac{1}{2} mc^2 - \frac{GMm}{r_{ML}} + \frac{\hbar c^3}{8GM} \\
&=& \frac{1}{2} mc^2 - \frac{GMm}{2GM/c^2} +  \frac{\hbar c^3}{8GM} \\
&=& \frac{\hbar c^3}{8GM}.
\end{eqnarray}
This energy corresponds to a temperature ($E/k_B$) of
\begin{equation}
T = \frac{\hbar c^3}{8GMk_B},
\end{equation}
exactly the same as the GR result for the same calculation and quite close to the Hawking temperature. We will call this temperature the Michell-Laplace-Hawking temperature $T_{MLH}$

We would like to calculate an entropy associated with this temperature. We start by assuming a small bit of matter of mass $dM$ falls from infinity into the M-L black hole, which means that its velocity as it crosses into the black hole is $c$.  The energy inside the black hole increases by $dMc^2/2$, and using the Clausius relation, $dE = T_{MLH} dS$, we have
\begin{equation}
\frac{1}{2}dM c^2 = \frac{\hbar c^3}{8GMk_B} dS,
\end{equation}
or
\begin{equation}
\frac{4Gk_B}{\hbar c}MdM = dS,
\end{equation}
and integrating,
\begin{equation}
S = \frac{2GM^2 k_B}{\hbar c}.
\end{equation}
In terms of the area of the black hole, $A = 4\pi r^2_{ML}$,
\begin{equation}
S = \frac{k_B A}{8\pi \ell^2_P},
\end{equation}
comparable to the Bekenstein-Hawking entropy $S = \frac{k_B A}{4\ell^2_P}$.

Note that all of these results are independent of the mass of the produced particle $m$, so {\it formally} we can take $m \rightarrow 0$ and model a photon. It is not unreasonable to assume that this ``photon'' would move with velocity $c$,
justifying our original assumption of the particle being produced with this velocity.

We need to point out a few facts.  One is that if we define the ``surface gravity'' $\kappa$ of the M-L black hole as the gravitational force per unit mass at $r_{ML}$ or the acceleration of a particle at the surface, 
\begin{equation}
\kappa = \frac{c^4}{4GM},
\end{equation}
and our Michell-Laplace temperature is
\begin{equation}
T_{ML} = \frac{\hbar \kappa}{2k_B c},
\end{equation}
equal to $\pi$ times the GR result.  It is easy to show that this is a universal relation for any Newtonian-type theory that gives a Newtonian potential $\Phi$ that is a function of radius only.  We have 
\begin{equation}
T = E_{\infty}/k_B = \frac{1}{k_B}\left [ \frac{1}{2}mc^2 + m\Phi(r_{ML} + l_c/2)\right],
\end{equation}
or to first order in $l_c$,
\begin{equation}
T = \frac{1}{k_B}\left [ \frac{1}{2} mc^2 +m\Phi(r_{ML}) + \frac{\hbar}{2c} \frac{d\Phi}{dr} \bigg |_{r = r_{ML}} \right ].
\end{equation} 
If $\Phi$ is a function of radius only, $\kappa = |\nabla \Phi (r_{ML})| = \frac{d\Phi}{dr}\bigg |_{r = r_{ML}}$, and
$\frac{1}{2} mc^2 + m\Phi(r_{ML}) = 0$,
\begin{equation}
T = \frac{\hbar\kappa}{2ck_B}.
\end{equation}
Of course non-Newtonian theories give more complicated $\Phi$'s, as we will see in Section V.

The Bekenstein entropy-area relation, $S=(k_B/4l^2_P) A$,  becomes, in the Newtonian case, $S = (k_B/8\pi l^2_P) A$.

\section{Newtonian field equations from entropy}

Jacobson and several coauthors \cite{jacprl},\cite{eling},\cite{gued}, \cite{jacmohd}, \cite{jacent} have considered the possibility of deriving the Einstein field equations from the Clausius relation, as well as several possible other relativistic field equations. We would like to consider the possibility of similar derivations for Newtonian physics and, in the next Section, modified Newtonian dynamics.

Jacobson, in Ref. \cite{jacprl}, considers the infall of matter through an Unruh horizon as a  touchstone of his derivation. We will consider the infall of matter through the sphere that is the boundary of a Michell-Laplace black hole of mass $M$. The Michell-Laplace radius is again $r_{ML} = 2GM/c^2$. We will assume that this black hole is surrounded by a radially infalling cloud of particles of constant density $\rho$. If we assume that the infall velocity of the cloud at $r_{ML}$ is $c$ (as if the particles were falling from infinity beginning with velocity zero).  The kinetic (heat) energy of the cloud at the horizon would be $\rho c^2/2$, and the simplest way of defining the amount of energy entering the black hole (of area $A$) per unit of time would be
\begin{equation}
\frac{dE}{dt} = \frac{1}{2} \rho c^3 A. \label{eeq}
\end{equation}
(notice that the energy \textit{loss} from the environment would be $-\rho c^2 A/2$). We will return to this equation later.

One of the bases for the Jacobson calculation is the Raychaudhuri equation. In elementary Newtonian theory no such equation exists.  However, we can find an equivalent equation by slightly modifying the Cartan formulation of Newtonian theory, that is, Cartan defined an affine (originally with no metric) space by writing the Newtonian equation of particle motion as a geodesic equation,
\begin{equation}
\frac{d^2x^{\mu}}{d\tau^2} + \Gamma^{\mu}_{\alpha \beta} \frac{dx^{\alpha}}{d\tau}\frac{dx^{\beta}}{d\tau} = 0,
\end{equation}
and defined $\Gamma^i_{00} = \Phi_{,i}$, $\Phi$ the Newtonian potential, and the rest of the $\Gamma^{\mu}_{\alpha \beta} = 0$. In this case $\tau = at + b$, and we can choose $b = 0$, $a = 1$, $\tau = t$ and
\begin{equation}
\frac{d^2 x^i}{dt^2} = -\Phi_{,i}. \label{newton}
\end{equation}
If we calculate $R_{\mu \nu}$, we find
\begin{equation}
R_{00} = \nabla^2 \Phi, \qquad {\rm rest \quad zero},
\end{equation}
and the field equation is
\begin{equation}
R_{00} = 4\pi G \rho.
\end{equation}
Deriving the Raychaudhuri equation for the Newton-Cartan theory needs a metric which does not exist for the standard theory. However, we can define a metric $g_{\mu \nu}$, but such a modified Newton-Cartan formulation will have to have nonmetricity, that is, the most general form of any connection coefficients \cite{Schout} is
\begin{equation}
\Gamma^{\sigma}_{\nu \alpha} = \frac{1}{2} g^{\sigma \mu} (g_{\mu \nu, \alpha} + g_{\mu \alpha,\nu} - g_{\alpha \nu, \mu}) - 
\frac{1}{2} (T_{\alpha}\,^{\sigma} \, _{\nu} + T_{\nu}\,^{\sigma} \, _{\alpha} - T_{\alpha \nu}\, \, ^{\sigma}) 
- \frac{1}{2} (Q_{\alpha}\,^{\sigma} \, _{\nu} + Q_{\nu}\,^{\sigma} \, _{\alpha} - Q ^{\sigma} \,_{\alpha \nu}),
\end{equation}
where $T^{\mu}_{\alpha \beta}$ is the torsion tensor and $Q_{\alpha \mu \nu} \equiv g_{\mu \nu; \alpha}$ is the nonmetricity tensor.  Here we want to have $g_{\mu \nu} =$ const., $T^{\sigma}_{\mu \nu} = 0$ and 
\begin{equation}
 Q_{0i0} = - \Phi_{,i},
\end{equation}
rest zero. We can take $g_{\mu \nu} = \delta_{\mu\nu}$ to give us Eq. (\ref{newton}). Notice that in order to keep a purely Newtonian theory we are not allowed to make coordinate transformations that change $t$.
 
The Raychaudhuri equation for a vector field $u^{\mu}$ can be derived (See, for example, Dadhich \cite{dady}) by defining the usual quantities,
\begin{equation}
\theta = u^{\mu}_{;\mu},
\end{equation}
\begin{equation}
\sigma_{\mu \nu} = u_{(\mu ; \nu)} - \frac{1}{3} \theta (g_{\mu \nu} - u_{\mu} u_{\nu}) -\dot u_{(\mu} u_{\nu )},
\end{equation}
($\dot u^{\mu} \equiv u^{\mu}_{;\nu} u^{\nu}$), and
\begin{equation}
\omega_{\mu \nu} = u_{[\mu;\nu]} - \dot u_{[\mu} u_{\nu]}.
\end{equation}
Even with nonmetricity the Riemann tensor is still defined by
\begin{equation}
u^{\mu}_{;\nu ; \gamma} - u^{\mu}_{;\gamma ; \nu} = -R^{\mu}_{\alpha \nu \gamma} u^{\alpha},
\end{equation}
and, putting $\mu = \nu$ and multiplying by $u^{\gamma}$, we find
\begin{equation}
\theta_{,\mu} u^{\mu} - u^{\mu}_{;\gamma ;\mu} u^{\gamma} = -R_{\mu \nu}u^{\mu} u^{\nu}. \label{thetaeq}
\end{equation}
The usual derivation of the Raychaudhuri equation rewrites the second term on the left-hand-side as $u_{\mu ; \nu} u^{\nu ; \mu}$ by using $(u^{\nu}_{;\mu} u^{\mu})_{;\nu} = u^{\nu}_{;\mu ;\nu} u^{\mu} + u^{\nu}_{;\mu} u^{\mu}_{;\nu}$ and the fact that the left-hand-side is zero for a geodesic field, and finally we have
\begin{equation}
u_{\mu ;\nu} u^{\nu ;\mu} = (\sigma_{\mu \nu} \sigma^{\mu \nu} - \omega_{\mu \nu}\omega^{\mu \nu} ) + \frac{1}{3} \theta^2,
\end{equation} 
so we arrive at
\begin{equation}
\frac{d\theta}{d\lambda} \equiv \theta_{,\mu} u^{\mu} = -(\sigma_{\mu \nu} \sigma^{\mu \nu} - \omega_{\mu \nu}\omega^{\mu \nu} ) -  \frac{1}{3} \theta^2  - R_{\mu \nu}u^{\mu} u^{\nu}.
\end{equation}
Note that these final steps rely heavily on being able to bring $g_{\mu \nu}$ inside the covariant derivative, i.e. $g_{\mu \nu ;\alpha} = 0 \Rightarrow  u^{\mu}_{;\nu}g_{\mu \alpha} = u_{\alpha ;\nu}$, but this is \textit{not} true for a space with nonmetricity.

With these cautions, as in the relativistic case we will be interested in a pencil of geodesics propagating radially from the horizon of the Michell-Laplace black hole.  We will assume that the tangent vector to this field is $u^{\mu} \equiv (u^0,\tilde u n^i)$, $n^i$ a constant radial normal to the black hole. The geodesic equation, $u^{\mu}_{;\alpha} u^{\alpha} = 0$ gives for $\mu = 0$,
\begin{eqnarray}
u^0_{;\alpha} u^{\alpha} &=& u^0_{,0} u^0 + u^0_{,i} u^i + \Gamma^0_{\mu \alpha} u^{\mu} u^{\alpha} \\
&=& u^0_{,0} u^0 + u^0_{,i} u^i = 0.
\end{eqnarray}
The standard solution is $u^0 =$ const. If this is so, $\theta$ becomes
\begin{equation}
u^0_{,0} + \tilde u_{,i} n^i + \Gamma^{\mu}_{\alpha \mu}u^{\alpha} = \tilde u_{,i}n^i.
\end{equation}
Using $\Gamma^{\mu}_{\alpha \mu}$ = 0 and (\ref{thetaeq}) and replacing $u^{\nu}_{;\mu ;\nu} u^{\mu}$ by $-u^{\nu}_{;\mu} u^{\mu}_{;\nu}$, we find
\begin{equation}
\frac{d\theta}{dt} + u^{\nu}_{;\mu} u^{\mu}_{;\nu} = -R_{\mu \nu}u^{\mu} u^{\nu}
\end{equation}
\begin{equation}
=\frac{d\theta}{dt} + \tilde u_{,i} \tilde u_{,j} n^i n^j = \frac{d\theta}{dt} + \theta^2 = -R_{00} (u^0)^2. \label{realthet}
\end{equation}
If, as in the relativistic case , we are interested in the evolution of $\theta$ over a short time, so we discard the $\theta^2$ term \cite{jacprl}, and we find
\begin{equation}
\frac{d\theta}{dt} = -R_{00} (u^0)^2 = -\nabla^2 \Phi (u^0)^2. 
\end{equation}
As in the relativistic case, we find
\begin{equation}
\frac{d\theta}{dt} = \frac{d}{dt}\left ( \frac{1}{A}\frac{dA}{dt}\right ) = -\nabla^2 \Phi(u^0)^2.
\end{equation}

Solving this equation for a short time interval, we have (expecting $\nabla^2 \phi$ to be constant since $\rho =$ const. and $u^0$ = const.)
\begin{equation}
\frac{1}{A} \frac{dA}{dt} = -t\nabla^2 \Phi (u^0)^2,
\end{equation}
and, if $A(t = 0) \equiv A_0$,
\begin{eqnarray}
A &=& A_0 e^{-\int t\nabla^2 \Phi (u^0)^2 dt} \\
&\approx& A_0 - A_0 \int t\nabla^2 \Phi (u^0)^2 dt = A_0 + \delta A.
\end{eqnarray}
We will now take the entropy $S$ to be $\eta A$,
\begin{equation}
S = \eta A_0 - \eta A_0 \int t\nabla^2 \Phi (u^0)^2 dt = S_0 + \delta S.
\end{equation}
Instead of using (\ref{eeq}) for $dE/dt$, for consistency we will follow Jacobson in writing
\begin{equation}
\frac{dE}{dt} = \frac{t}{2t_{(0)}} \rho c^3 A, \label{neweeq}
\end{equation}
where he takes the Unruh value, $t_{(0)} = c/\kappa$.  Now
\begin{equation}
E = E_0 - A_0 \int \frac{\kappa}{2c} t \rho c^3 dt = E_0 + \delta E,
\end{equation}
which, with (taking $u^0 =1$)
\begin{equation}
 T_{ML} \delta S = -\frac{\hbar \kappa}{2k_B c} A_0 \eta \int \nabla^2 \Phi t dt,
\end{equation}
and
\begin{equation}
\delta E = -\frac{\kappa}{2} c^2 A_0\int \rho tdt,
\end{equation}
and the Clausius relation
\begin{equation}
\delta E = T_{MLH} \delta S
\end{equation}
implies, if we want $\nabla^2 \Phi = 4\pi G\rho$,
\begin{equation}
\eta = \frac{k_B c^3}{4\pi \hbar G} = \frac{k_B}{4\pi l^2_P},
\end{equation}
that is, $1/\pi$ times the usual $\eta$, $\eta = 1/4 l^2_P$ and twice the Newtonian $\eta$ from Sec. II.

\section{Newtonian calculation of the ``Hawking'' temperature and entropy for an $f(R)$ theory}

We want to consider similar calculations for a simple example of more complicated Newtonian theory. Following an article by Quandt and Schmidt \cite{QS}, who found such theories using the equations from a Lagrangian of the form 
\begin{equation}
{\cal L} = \left(R + R\sum^p_{k = 0} a_k \Box^k R \right)\sqrt{-g} ,
\end{equation}
where $\Box$ is the d'Alembertian ($R^{; \mu}_{\, ;\mu})$.

In our case we will consider a higher curvature theory based on the above Lagrangian for $p = 0$ and with $a_0 \equiv \ell_0^2$ for dimensional reasons, 
\begin{equation}
{\cal L}  = (R + \ell^2_0 R^2) \sqrt{-g},
\end{equation}
the field equations for a Lagrangian of this form are
\begin{equation}
G_{\mu \nu} + \ell_0^2 R(2R_{\mu \nu} - \frac{1}{2}g_{\mu \nu} R) + 2\ell_0^2 g_{\mu 
\nu}\Box R - 2\ell_0^2 R_{;\mu ;\nu} =\frac{8\pi G}{c^4}T_{\mu \nu}.
\end{equation}

If we now substitute the linearized version of a spherically symmetric metric in isotropic spherical coordinates \cite{QS},
\begin{equation}
ds^2 = -\left [1- 2\varepsilon \frac{\phi(r)}{c^2} \right ] dt^2 + \left [1 + 2\varepsilon \frac{\psi(r)}{c^2} \right ] (dr^2 + r^2d\theta^2 + r^2 \sin^2 \theta d\varphi^2),\label{metric}
\end{equation}
and since $R_{\mu \nu}$ and $R$ are both of order $\varepsilon$, the above equation to order $\varepsilon$ becomes
\begin{equation}
G_{\mu \nu} + 2\ell^2_0 g_{\mu \nu}\Box R - 2\ell^2_0 R_{;\mu ;\nu} = \frac{8\pi G}{c^4} T_{\mu \nu}.
\end{equation}
The trace of this equation is
\begin{equation}
-R + 6\ell^2_0 \Box R = \frac{8\pi G}{c^4}T. \label{trace}
\end{equation}
The final field equations are diagonal, and
\begin{equation}
G_{00} - 2\ell_0^2 \Box R = \frac{8\pi G}{c^4} T_{00},
\end{equation}
\begin{equation}
G_{rr} + 2\ell^2_0 \Box R - 2\ell^2_0 R^{\prime \prime} = \frac{8\pi G}{c^4} T_{rr},
\end{equation}
\begin{equation}
\frac{1}{r^2}G_{\theta \theta} + 2\ell_0^2 \Box R - 2\ell^2_0 \frac{1}{r} R^{\prime} = \frac{8\pi G}{r^2 c^4}T_{\theta \theta}.
\end{equation}
(where $\prime$ is $d/dr$) and the left-hand-side of the ($\varphi \varphi$) equation divided by $\sin^2 \theta$ is the same as that of the ($\theta \theta$) equation. We also have
\begin{equation}
R = \frac{2}{c^2}(\nabla^2 \phi -2 \nabla^2 \psi).\label{finR}
\end{equation}
For the metric (\ref{metric}) and following Quandt and Schmidt in using $T_{\mu \nu} = \rho c^2 \delta^0_{\mu} \delta^0_{\nu}$ we find (for this metric $\Box$  becomes the flat space $\nabla^2$),
\begin{equation}
(00):\,\,-\frac{2}{c^2} \nabla^2 \psi - 2\ell^2_0 \nabla^2 R = \frac{8\pi G}{c^2} \rho, \label{00}
\end{equation}
\begin{equation}
(rr):\,\, \frac{2}{r c^2}(\psi^{\prime} - \phi^{\prime}) + 2\ell^2_0 \nabla^2 R - 2\ell^2_0 R^{\prime \prime} = 0,
\end{equation}
\begin{equation}
(\theta \theta \, \& \, \varphi \varphi/\sin^2 \theta):\,\,-\frac{1}{c^2} \nabla^2 \phi + \frac{1}{c^2} \nabla^2 \psi + \frac{1}{c^2}(\phi^{\prime} - \psi^{\prime}) + 2\ell^2_0 \nabla^2 R - 2\frac{\ell^2_0}{r} R^{\prime} = 0.
\end{equation}
The trace equation is now
\begin{equation}
-R + 6\ell^2_0 \nabla^2 R = \frac{8\pi G}{c^2}\rho,
\end{equation}
which can be solved for $\nabla^2 R$ as
\begin{equation}
\nabla^2 R = \frac{1}{6\ell^2_0} \left ( -\frac{8\pi G}{c^2} \rho + R \right ),
\end{equation}
and plugging this into Eq.(\ref{00}),
\begin{equation}
-\frac{2}{c^2}\nabla^2 \psi + \frac{8\pi G}{c^2}\rho - \frac{1}{3}R = \frac{8\pi G}{3c^2} \rho.
\end{equation}
Using $R = (2/c^2)(\nabla^2\phi - 2\nabla^2\psi)$, we can eliminate $\nabla^2 \psi$ using (\ref{finR}) and
\begin{equation}
\nabla^2 \psi = -\nabla^2 \phi - 8\pi G \rho.
\end{equation}
Finally, eliminating $\psi$ from $R$ and $G_{00}$, Eq. (\ref{00}) becomes
\begin{equation}
\nabla^2 \phi - 6\ell^2_0 \nabla^2(\nabla^2 \phi) - 32\pi G \ell_0^2 \nabla^2 \rho = -4\pi G \rho. \label{fldeq}
\end{equation}
Notice that linearization has reduced a higher curvature equation to a {\it higher-derivative} equation. Also,
an important fact is that $\nabla^2 \rho$ appears.

As in \cite{QS}, we take $\rho$ to be that of a point mass $M$, $\rho = M\delta({\bf r})$, and substitute 
\begin{equation}
\phi = \frac{A}{r} + \frac{B}{r} e^{-r/\sqrt{6}\ell_0},
\end{equation}
and find (treating the second term as a distribution where $\nabla^2 [((B/r)e^{-r/\sqrt{6}\ell_0})]$ is $(B/6\ell^2_0 r)e^{-r/\sqrt{6}\ell_0} - 4\pi \delta ({\bf r})$),
\begin{equation}
A\delta({\bf r}) - 6\ell_0^2[A + B - \frac{4}{3} GM]\nabla^2 \delta({\bf r}) = GM\delta ({\bf r})).
\end{equation}
and putting the coefficients of $\delta ({\bf r})$ and $\nabla^2 \delta ({\bf r})$ separately equal to zero, we find
\begin{equation}
A = GM, \qquad B = \frac{1}{3} GM.
\end{equation}
The Newtonian potential $\Phi$ is $-\phi$.

At this point we will take $\sqrt{6} \ell_0$ to be the Planck length $\ell_P$ for reasons that will become clear below, and $\Phi$ is
\begin{equation}
\Phi = - \frac{GM}{r} \left ( 1 + \frac{1}{3} e^{-r/\ell_P} \right ). \label{modpot}
\end{equation}

Notice that for radii reasonably larger than the Planck length the potential has the form
\begin{equation}
\Phi = -\frac{GM}{r} [1 + \Phi_1 (r)], \label{abpot}
\end{equation}
Where $\Phi_1$ is small.  Both here and in our companion article we are interested in potentials of this form. In the case of the companion article we investigate quantum corrections to forces and potentials derived from generalized entropies by using the techniques of Ref. \cite{MOR}. In those cases, the Planck length appears naturally from the use of the relation between black hole entropy and area, $S_{BH} = A/4\ell_P$, so we will make connection with that work by assuming that (\ref{modpot}) comes from (\ref{fldeq}) where the higher order terms come from an as yet unknown quantum gravity, justifying our assumption that $\ell_P = \sqrt{6}\ell_0$.

This said, we can study a potential of the form of (\ref{abpot}).  For such a potential we can easily find the Michell-Laplace radius, $r_{ML}$, for a black hole by assuming zero velocity at infinity for a particle leaving $r_{ML}$
with velocity $c$, and, if we assume that $r_{ML}$ is a black-hole radius of a reasonably large mass, $\Phi_1$ should be very small if $\Phi_1$ is a function of $r/\ell_P$, so we have
\begin{equation}
\frac{1}{2}c^2 - \frac{GM}{r_{ML}}[1 + \Phi_1 (r_{ML})] = 0,
\end{equation}
and solving to first order in $\Phi_1$,
\begin{equation}
\frac{2GM}{c^2} = r_{ML}[1 - \Phi_1 (r_{ML})].
\end{equation}

Since the correction $\Phi_1 (r_{ML})$ is very small, we assume that $r_{ML} = 2GM/c^2 + \lambda$, $\lambda$ small.  Expanding $\Phi_1$ to first order in $\lambda$,
\begin{equation}
\frac{2GM}{c^2} \Phi_1 (2GM/c^2) = \lambda \left (1 + \Phi_1 (2GM/c^2) + M\frac{d\Phi_1}{dM}(2GM/c^2) \right).
\end{equation}
Dividing by $1 + \Phi_1 + M(d\Phi_1/dM)$ and assuming that $Md\Phi_1/dM$ is also small, we have
\begin{equation}
r_{ML} \cong \frac{2GM}{c^2} \left [1 + \Phi_1 \left (\frac{2GM}{c^2} \right ) \right ].
\end{equation}
As before, if a particle of mass $m$ leaves $r_{ML} + \ell_c/2$ with velocity $c$, and we expand $\Phi (r_{ML} + \ell_c/2)$ to first order in $\ell_c$, and using the fact that $m\ell_c = \hbar/c$, the energy at infinity is
\begin{equation}
E_{\infty} = \frac{1}{2} Mc^2 \left (\frac{\ell_P}{r_{ML}} \right )^2 \left [ 1 + \Phi_1 (r_{ML}) - r_{ML} \frac{d\Phi_1}{dr_{ML}} (r_{ML}) \right ] = k_B T_{MLH}.
\end{equation}
Using $M = (c^2/2G)r_{ML} [1 - \Phi_1 (r_{ML})]$,
\begin{equation}
T_{MLH} = \frac{1}{2k_B} M_P c^2 \left( \frac{\ell_P}{r_{ML}} \right ) \left [1 - r_{ML} \frac{d\Phi_1}{dr_{ML}} (r_{ML}) \right ].
\end{equation}

It will be useful later to use $T_{MLH}$ as a function of $M$, and inserting $r_{ML}$ as a function of $M$ and expanding to first order in $\Phi_1$ and $d\Phi_1/dM$,
\begin{equation}
T_{MLH} = \frac{M_P^2 c^2}{8k_B M}\left [ 1 - \Phi_1 (2GM/c^2)  - M\frac{d\Phi_1}{dM} (2GM/c^2) \right ]. \label{TMLHM}
\end{equation}

We can now find the equivalent of the Bekenstein-Hawking entropy using
\begin{equation}
dS = k_B \frac{dM c^2}{2T_{MLH}}. \label{dStemp}
\end{equation}
We have, using $dM = [dM(r_{ML})/dr_{ML}]dr_{ML}$,
\begin{equation}
\frac{dS}{dr_{ML}} = k_B\frac{r_{ML}}{\ell_P^2}[1 - \Phi_1(r_{ML})], \label{dSdr}
\end{equation}
or
\begin{equation}
S = \frac{k_B}{2} \left [ \frac{r_{ML}^2}{\ell_P^2} - 2\int\frac{r_{ML}}{\ell_P^2} \Phi_1 dr_{ML} \right].
\end{equation}
Using ({\ref{dStemp}) with (\ref{TMLHM}), $S$ as a function of $M$ is (integrating by parts),
\begin{equation}
S = 2k_B\left (\frac{M}{M_P} \right ) \left [ 1 + 2\Phi_1 (2GM/c^2) - \frac{2}{M_P^2} \int M\Phi_1 (2GM/c^2) dM \right ].
\end{equation}

For our $\Phi$, $\Phi_1 = \frac{1}{3} e^{-r/\ell_P}$, and
\begin{equation}
r_{ML} \cong \frac{2GM}{c^2} \left[ 1 + \frac{1}{3}e^{-2M/M_P} \right],
\end{equation}
and
\begin{eqnarray}
T_{MLH} &=& \frac{1}{2k_B} M_P c^2 \left ( \frac{\ell_P}{r_{ML}} \right )^2 \left [ 1 + \frac{r_{ML}}{\ell_P} e^{-2r_{ML}/\ell_P} \right ] \\
&=& \frac{M_P^2 c^2}{8k_B M} \left [ 1 - \frac{1}{3} e^{-2M/M_P} + \frac{2}{3} \frac{M}{M_P} e^{-2M/M_P} \right ].
\end{eqnarray}

Finally, integrating (\ref{dStemp}),
\begin{eqnarray}
S &=& \frac{k_b}{2} \left [ \left (\frac{r_{ML}}{\ell_P}\right )^2 + \frac{2}{3} \left ( \frac{r_{ML}}{\ell_P} + 1 \right ) e^{-r_{ML}/\ell_P} \right ]\\
&=& 2k_B \left ( \frac{M}{M_P} \right ) \left [ 1 + \left (\frac{5}{6} + \frac{M}{3M_P} \right) e^{-2M/M_P} \right ].
\end{eqnarray}

We now need the entropy-area relation.  The area of our black hole is $A = 4\pi r_{ML}^2$, and from Eq.(\ref{dSdr}), $S$ as a function of $A$ becomes
\begin{equation}
dS = \frac{k_B}{8\pi \ell^2_P}\left [ 1 - \frac{1}{3}e^{-\sqrt{A}/2\sqrt{\pi} \ell_P} \right ] dA. \label{dSA}
\end{equation}
This gives
\begin{equation}
S = \frac{k_B}{8\pi \ell_P^2}  \left [ A - \frac{8\pi}{3}\ell_P^2  e^{-\sqrt{A}/2\sqrt{\pi}\ell_P} -\frac{4\sqrt{\pi}}{3} \ell_P \sqrt{A} e^{-\sqrt{A}/2\sqrt{\pi} \ell_P} \right ],
\end{equation}
a much more complicated function of $A$ than in the Newtonian case.

\section{Conclusions and suggestions for further research}

We have shown how Newtonian and modified Newtonian gravity with the addition of some extremely simplified ideas from quantum field theory can reproduce results from general relativity, at least at the level of back-of-the-envelope calculations.  These included a simplified Hawking temperature, the Michell-Laplace temperature, an entropy similar to the Bekenstein result, a Newtonian version of Jacobson's calculation of field equations from entropy, and an investigation of the ``Hawking'' temperature and entropy for a higher-order modified Newtonian theory.

On thing that has not been done is to try to deduce field equations for this modified theory directly from the entropy as a function of area.  Calculations similar to those of Refs. \cite{gued} and \cite{eling} for higher-order relativistic theories of gravity could be extended to modified Newtonian theories (MoND's).

Another idea that will be discussed in the companion article \cite{MOR} is an attempt to find field equations for modified Newtonian theories directly from generalized entropies that can be calculated by means of statistical  interpretations of entropy or, for example, from loop quantum gravity. These entropies all have the property of being functions of the Boltzmann entropy $S_B$.

As a preliminary calculation we can use the entropy of the modified Newtonian theory of Sec. IV.  In Ref. \cite{OO} an idea due to Verlinde \cite{verl} and extended to give a more generalized Newtonian picture \cite{mod} is used to modified Newtonian potentials arising from generalized entropies that depend only on the probability.  The main idea is to assume that the area of a black hole is proportional to the Boltzmann entropy $S_B$, $\alpha A = S_B$, with $S = S(S_B)$ for a generalized entropy.

For our Newtonian calculation we have to take the Boltzmann entropy of our theory to be
\begin{equation}
S_B = \frac{k_B A}{8\pi \ell^2_P},
\end{equation}
and the generalized entropy associated with the theory discussed in Sec. IV will be
\begin{equation}
S = S_B  - \frac{1}{3} e^{-\sqrt{2S_B/k_B}} - \frac{\sqrt{S_B}}{6\sqrt{\pi} \ell_P} e^{-\sqrt{2S_B/k_B}}. \label{dSB}
\end{equation}

In Refs. \cite{OO} and \cite{mod} a procedure to calculate the Newtonian (or modified Newtonian) force on a particle of mass $m$ was given. They write the Newtonian force due to a point mass $M$ as an entropic force (assuming a strictly radial force depending only on the radius $r$) as
\begin{equation}
{\bf F} = -\frac{GMm}{r^2} \frac{4\ell^2_P}{k_B} \frac{\partial S}{\partial A} \bigg |_{A = 4\pi r^2} \hat {\bf r},
\end{equation}
where they have used the relativistic relation $S = k_B A/4\ell_P^2$ rather than our Newtonian one.  In our case
we should write
\begin{equation}
{\bf F} = -\frac{GMm}{r^2} \frac{8\pi \ell_P^2}{k_B} \frac{\partial S}{\partial A}\bigg |_{A = 4\pi r^2} \hat {\bf r}. \label{enf}
\end{equation}

Using $dS/dA$ from (\ref{dSA}) the entropic force (\ref{enf}) becomes
\begin{equation}
\mathbf{F} = -\frac{GMm}{r^2}\left [1 - \frac{1}{3} e^{-r/\ell_P} \right ]\hat {\bf r}. \label{enf2}
\end{equation}
For our potential (\ref{modpot}) the force, $-\nabla \Phi \hat {\bf r}$, is
\begin{equation}
{\bf F} = \frac{-GMm}{r^2} \left [ 1 + \frac{1}{3} e^{-r/\ell_P} + \frac{1}{3} \left ( \frac{r}{\ell_P} \right ) e^{-r/\ell_P} \right ] \hat {\bf r}. \label{realforce}
\end{equation}
The entropic force is far from this force, so the circle does not close.  However, the entropic force {\it does} generate another modified Newtonian theory of gravity.  We can find the potential $\Phi_E$ that generates this force,
$-\int ({\bf F} \cdot \hat {\bf r}) dr$, and
\begin{equation}
\Phi_E = -\frac{GMm}{r}\left [1 - \frac{1}{3}e^{-r/\ell_P} + \frac{r}{\ell_P} \eint (r/\ell_P) \right ],
\end{equation}
$\eint$ an exponential integral, a somewhat more unusual potential.

Another way of calculating the force would be to define what could be called a ``Schwarzschild area,'' $A_S$, $A_S = 16\pi G^2 M^2/c^4$, or $M = \sqrt{A_S} c^2/4\sqrt{\pi} G$. This gives us
\begin{equation}
\frac{dS}{dA_S} = \frac{1}{8\pi \ell_P^2} \left [1 + \Phi_1 (\sqrt{A_S}/2\sqrt{\pi}\ell_P) + \sqrt{A_S}\frac{d\Phi_1}{d\sqrt{A_S}} (\sqrt{A_S}/2\sqrt{\pi} \ell_P) \right ],
\end{equation}
or in our case,
\begin{equation}
\frac{dS}{dA_S} = \frac{1}{8\pi \ell_P^2} \left [ 1 + \frac{1}{3} e^{-\sqrt{A_S}/2\sqrt{\pi} \ell_P} - \frac{\sqrt{A_S}}{6\sqrt{\pi} \ell_P} e^{-\sqrt{A_S}/2\sqrt{\pi} \ell_P}\right ].
\end{equation}
and the entropic force is
\begin{equation}
\mathbf{F} = -\frac{GMm}{r^2} \left [ 1 + \frac{1}{3} e^{-r/\ell_P} - \frac{1}{3} \left(\frac{r}{\ell_P}\right) e^{-r/\ell_P} \right ],
\end{equation}
which is closer to $\-m \mathbf{\nabla} \Phi$, except for the sign of the last term in brackets. The major difficulty, however, is that $A_S$ has nothing to do with the area of the black hole.

One question we have not addressed is what field equation would generate the $\Phi$ given above.  This is difficult to answer.  It is similar to asking what ODE has as its solution the algebraic function $y = x^3$.  Without any restrictions on our differential equation there are an infinite number of answers.  In our companion article \cite{MOR} we discuss this problem.

One possible solution would be to use a Jacobson-type analysis for higher-curvature theories mentioned above. For our simplified Newtonian calculation in Sec. III it would be difficult to reconcile Eq. (\ref{eeq}) that is linear in $A$ with an entropy $S(A)$ that is a more complicated function of $A$.

Another possible avenue to explore is Wald's derivation of black hole entropy as a Noether charge \cite{Wald, Iyer}.
There are a number of articles that attempt to find such an entropy for $f(R)$ theories of gravity with varying levels of success. We plan to consider this possibility in our Newtonian context where explicit black hole solutions are more easily found, perhaps from the viewpoint of the metric-affine formulation of Section III.   

\begin{acknowledgments}
The work of A.M.M. is supported by a CONACyT post-doctoral grant. O.O. was supported by CONACyT Project numbers 257919 and 258982, PROMEP and UG Projects.
\end{acknowledgments}

\appendix*
\section{Simplified electromagnetic and gravitational particle production}

Particle production by an electric field can be modeled by assuming that virtual particles 
appear and disappear inside a distance of the Compton wavelength of that particle. 
That is, we take the following Feynman diagram seriously.
\begin{figure}[h]
\centering
		\includegraphics[width=8cm,height=5cm]{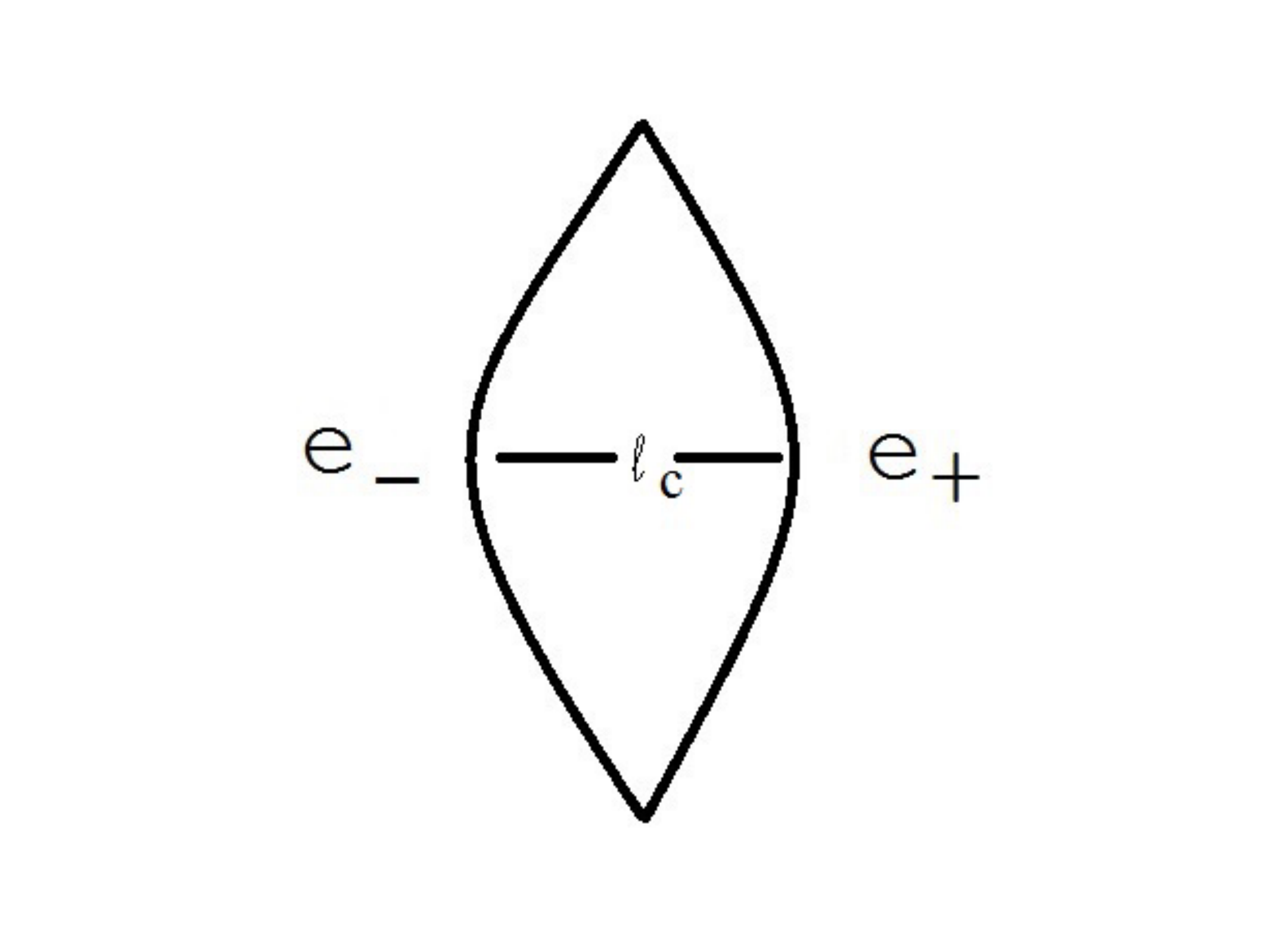}
	\caption{Feynman diagram for the virtual production of an electron-positron pair.}
	\label{fig:fdiag2}
\end{figure}
The two particles have a maximum separation of the Compton wavelength, $\ell_c = \hbar/mc$, where for an electron we use $m = m_e$. 

At the maximum separation the force of attraction between the two particles is $e^2/\ell_c^2$. If we apply a constant electric field, $E$, it causes a force separating the two particles of $eE$. The electric field will pull the virtual particles apart and make a real electron-positron pair when $E = e/\ell_c^2$, or
\begin{equation}
E = \frac{em^2 c^2}{\hbar^2}.
\end{equation}
For an electron this is 
\begin{equation}
E \approx 10^{18} \frac{statV}{cm} = 10^{17} \frac{V}{m}.
\end{equation} 
A typical Van de Graff gives an $E$ of a few tens of millions of volts per meter, $E \approx 10^6 V/m$, or eleven orders of magnitude less than what is needed to produce these particles from vacuum. 

Another possibility is to assume that we need enough electric field energy in a sphere of radius $\ell_c$ to be equal to twice the rest energy of a particle of mass $m$, $2mc^2$ \cite{FMHV}, or 
\begin{equation}
\left (\frac{E^2}{8\pi}\right) \left( \frac{4\pi}{3}\ell_c^3\right ) = 2mc^2,
\end{equation} 
or 
\begin{equation}
E^2 = \frac{12c\hbar}{\ell_c^4} = \frac{12e^2}{\alpha \ell_c^4}, 
\end{equation}
$\alpha$ the fine structure constant, or
\begin{equation}
E \approx \frac{40e}{\ell_c^2},
\end{equation}
about an order of magnitude more than above.  

If we attempt to have a gravitational field produce particles by a similar mechanism, we can consider the following ``Feynman'' diagram with a large mass $M$ a distance $R$ from a virtual pair. We consider the virtual pair to be two masses $m$. There is no need for a negative mass, since there is no conservation of mass (however, see the body of the paper for a caveat). Figure ~\ref{fig:gravpart2} shows this situation.
\begin{figure}[h]
\centering
		\includegraphics[width=8cm,height=5cm]{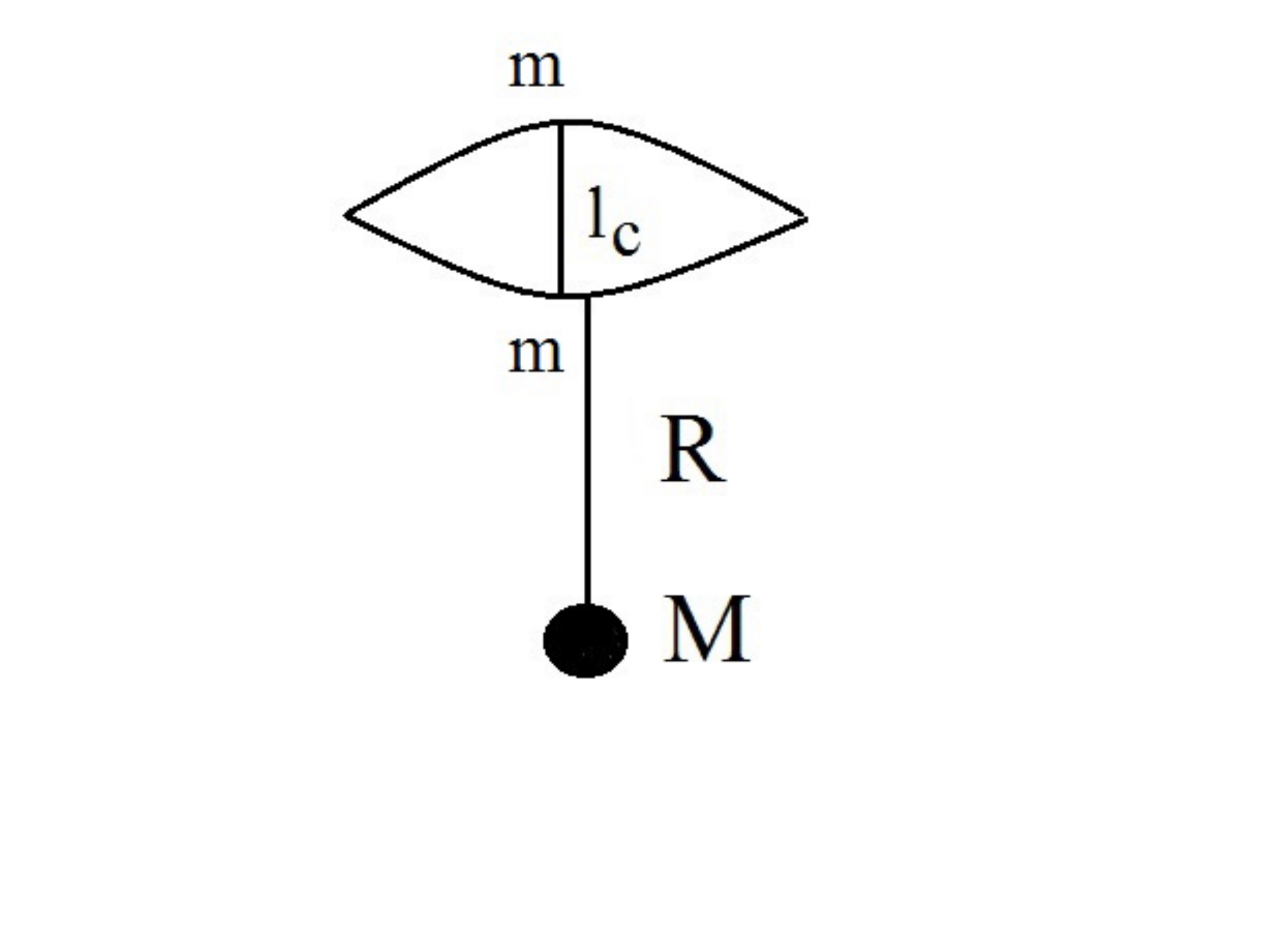}
	\caption{``Feynman'' diagram for the virtual production of two particle of mass $m$.}
	 \label{fig:gravpart2}
\end{figure}
For simplicity we will consider simply Newtonian gravity, the tidal force separating the two masses is
\begin{equation}
\frac{2GMm}{R^3}\ell_c.
\end{equation}
For this to overcome the gravitational attraction between the two particles we would need 
\begin{equation}
\frac{2GMm}{R^3}\ell_c = \frac{Gm^2}{\ell_c^2},
\end{equation}
or
\begin{equation}
\frac{M}{R^3} = \frac{m}{2\ell_c^3}.
\end{equation}
To produce particles of mass $m$ just at the point where the escape velocity is $c$ (supposing the particles separate at the speed of light) for the particles to escape to infinity, we need $R =2GM/c^2$, or a mass of 
\begin{equation}
M^2 = \frac{c^3 \hbar^3}{4G^3 m^4} = \frac{1}{4} \frac{M_P^6}{m^4},
\end{equation}
($M_P$ the Planck mass) or 
\begin{equation}
M = \frac{M_P^3}{m^2}.
\end{equation}
To produce particles with the mass of an electron, 
\begin{equation}
M \approx 10^{39}{\rm gm} \approx 10^6 M_\odot,
\end{equation}
more or less the mass of a globular cluster. For many years it was assumed that tidal forces would drive particle production, requiring gravitational fields of this magnitude. In GR this would involve $R_{\alpha \beta \mu \nu}$. 

In GR we can consider the same virtual particles in a spherical shell between the Schwarzschild radius of a black hole, $r_S$, and $r_S + \ell_c$ (see Fig. \ref{fig:shell3}). 
\begin{figure}[h]
	\centering
		\includegraphics[width=8cm,height=5cm]{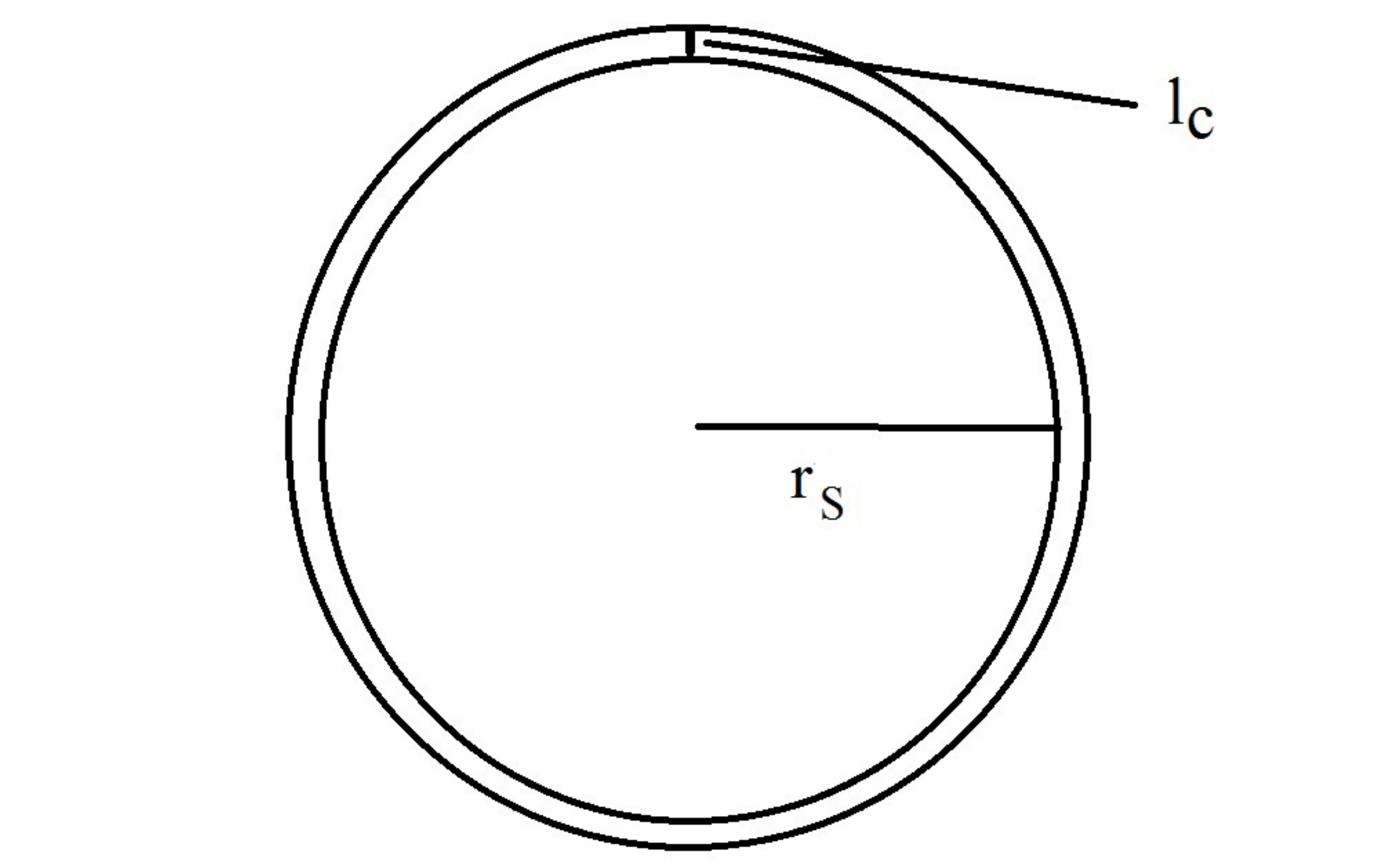}
	\caption{Shell of width $\ell_c$ around a black hole.}
		\label{fig:shell3}
\end{figure}
For radial particle motion in a Schwarzschild field we have
\begin{equation}
\frac{dr}{d\tau} = \sqrt{\varepsilon^2 - 1 + \frac{r_S}{r}},
\end{equation}  
where $\varepsilon$ is a dimensionless energy parameter. Assume that one member of the particle pair falls into the black hole and disappears, and the other escapes to infinity from $r_0 = r_S + \ell_c/2$ with velocity $c$. For $r = r_0$, 
\begin{equation}
\frac{dr}{d\tau}\bigg|_{r_0} = 1 \approx
\sqrt{\varepsilon^2 - \frac{\ell_c}{2r_S}},
\end{equation} 
and at infinity, 
\begin{equation}
\frac{dr}{d\tau} \bigg|_{r = \infty} = \sqrt{\varepsilon^2 - 1}.
\end{equation}
Solving, we find the kinetic energy at infinity $mc^2(\varepsilon - 1)$, using 
\begin{equation}
\varepsilon^2 = 1 + \frac{\ell_c}{2r_S},
\end{equation}
or 
\begin{equation}
\varepsilon \approx 1 + \frac{\ell_c}{4r_S},
\end{equation} 
so 
\begin{equation}
mc^2(\varepsilon - 1) = \frac{mc^2 \ell_c}{4r_S} = \frac{\hbar c^3}{8MG}.
\end{equation} 
Note that $m$ does not appear, so at least formally, this result applies to radiation as well as particles. The temperature associated with this energy is
\begin{equation}
T = \frac{\hbar c^3}{8MGk_B} = T_{MLH},
\end{equation} 
essentially the Hawking temperature, $T_H = \hbar c^3/8\pi GMk_B$. The factor of $\pi$ may be geometric.



\end{document}